# Four switching categories for thin-film and bulk ferroelectrics

X.J. Lou[*]

Department of Materials Science and Engineering, National University of Singapore, 117574, Singapore

**Abstract:**

We classify the switching kinetics of ferroelectrics including both epitaxial/polycrystalline thin films and single-crystalline/ceramic bulks at various applied fields into four categories, depending on whether the depolarization field and/or the polarization reversal induced by the switching promotion effect between adjacent parts can be neglected. We show that our statistical model developed very recently [Journal of Physics: Condensed Matter **21**, 012207 (2009)] in its generalized form applies to all these four categories. Finally we make the comparison between our model and the conventional Kolmogorov-Avrami-Ishibashi model and discuss the behavior of the switching currents for different *n*.

The traditional approach used to describe the switching kinetics of ferroelectrics is the Kolmogorov-Avrami-Ishibashi (KAI) model [1], based on the classical theory of Kolmogorov [2] and Avrami [3]. By assuming unrestricted domain growth in an infinite crystal, this model predicts the polarization change $\Delta P(t)$ to be:

$$\frac{\Delta P(t)}{2P_s} = 1 - \exp\left[-(t/\tau)^n\right] \qquad (1)$$

where *n* is the effective dimensionality and $\tau$ the characteristic time. This model could be further classified into two cases [1]: the *α* model, where nucleation occurs with a constant rate during switching (*n=D+1*), and the *β* model, where nucleation occurs only at *t=0* (*n=D*). *D* is the geometrical growth dimensionality of the system ($D=1$ for stripelike domains, $D=2$ for circular domains and $D=3$ for spherical domains).

---

[*] Correspondence email: mselx@nus.edu.sg

Although the KAI model gives a good description of the switching kinetics of ferroelectric single crystals [4] and sometimes epitaxial thin films [5], it encounters problems when it comes to correctly describing the domain reversal behavior of polycrystalline thin films [6-8]. The strong retardation of the switching curves at medium or low fields for these films has been explained via polarization processes with a broad distribution of relaxation times [6], a nucleation-limited-switching (NLS) model [8], or the Lorentzian distribution of logarithmic switching times [7].

Very recently, we showed that the polarization switching process in polycrystalline ferroelectric thin films or ceramic bulks can be accounted for using a statistical model incorporating a time-dependent depolarization field [9]. In this model, we assume that polarization reversal in a ferroelectric capacitor takes place in a *part-by-part* or *region-by-region* manner due to the *blocking* effect of grain boundaries [10], defect planes/dislocations and/or 90 ° domain walls [11]. Although it is a good approximation for ferroelectric polycrystalline thin films [10-12] and ceramics, this model needs some modifications or generalizations to describe the switching kinetics of grain-boundary/90 ° domain-wall-free single-crystalline materials and epitaxial thin films, where the Polarization Reversal induced by the Switching Promotion effect between Adjacent Parts (PRSPAP) is believed to be an important factor influencing their switching kinetics [by PRSPAP, we mean the polarization reversal induced by the switching promotion effect of the switched parts on an adjacent non-switched part via Domain-Wall Geometric Nucleation (DWGN) effect [13] under an applied field]. Indeed, Domain Wall Motion (DWM) [or PRSPAP in the present context] to some extent has long been observed by optical microscope or Scanning Probe Microscope in both single-crystalline bulk ferroelectrics [14, 15] and epitaxial ferroelectric thin films [16-18].

Here we propose that the switching kinetics of ferroelectrics including both epitaxial/polycrystalline thin films and single-crystalline/ceramic bulks at various applied fields $E_{appl}$ can be classified into four categories (Table I), depending on whether or not the depolarization field $E_{dep}$ and/or the effect of PRSPAP can be neglected. We show that our switching model [9] in its generalized form applies to all these four categories (For Category III and IV we will take the effect of PRSPAP into consideration).

**Category I:** polycrystalline thin films (or relatively thicker epitaxial films containing *blocking* 90 ° domain walls) at medium or low $E_{appl}$ [$n_{app}$<1] and **Category II**: polycrystalline thin films (or relatively thicker epitaxial films containing *blocking* 90 ° domain walls) at high $E_{appl}$ or polycrystalline (ceramic) bulks [$n_{app}$~1] ($n_{app}$ denotes "apparent" *n*)

These two categories have been investigated fully in our previous work [9]. Note that our model implies that relatively thicker epitaxial films containing *blocking* 90 ° domain walls also fall into Category I or II (Table I), depending on the field regime we are working on: in polycrystalline thin films (or ceramics) PRSPAP is inhibited mainly by grain boundaries containing planes of defects, while in relatively thicker epitaxial films this motion is blocked primarily by 90 ° domain walls. Indeed, recent work by Li and Alexe [11] shows that relatively high density of 90 ° domain walls appears in their epitaxial films with *d*>215 nm. They found that the switching kinetics of these relatively thicker films shows different features from those in the 90 ° domain-wall-free thinner films (see Fig 1-3 in Ref [11]), but is similar to those observed in polycrystalline $Pb(Zr,Ti)O_3$ (PZT) films [7]: e.g. polarization reversal is highly retarded at later switching stage and the KAI fitting gives rise to $n_{app}$<1 at medium or low $E_{appl}$ [7, 11].

Note that in our previous work [9] we ignored the effect of PRSPAP for Category I and II, which is obviously an *ideal* assumption. In reality, PRSPAP for the samples in Category I or II, though difficult, is possible, especially at high driving $E_{appl}$. And that is might be the main reason why $n_{app}$ appears to be slightly larger than one at high fields [9] (also see Fig 1 and 2 in Ref [7] for polycrystalline films and Fig 2e and Fig 3c in Ref [11] for epitaxial films containing a high density of 90 ° domain walls). Other reasons that are extrinsic and may also gives rise to $n_{app}$>1 are the problem of $t_w$ discussed in Ref [9] and that of the pulse rise time discussed later; these effects either screen the upper and lower parts of the "∫"-shaped switching curve or make it span less time decades than it is supposed to do (i.e., a larger $n_{app}$, possibly larger than 1).

For Category III and IV treated in the following sections, we, again, divide the capacitor *uniformly* into $M_0$ parts and view that under the total field (i.e., the sum of $E_{appl}$ and $E_{dep}$) polarization switching takes place in a *part-by-part* or *region-by-region* manner. However, unlike the assumption we made previously, the PRSPAP is taken into consideration now (Fig 1), consistent with the observations in both single-crystal bulks [14, 15] and epitaxial thin films [16, 19].

**Category III:** 90 ° domain-wall-free single-crystalline (epitaxial) thin films at high $E_{appl}$ or single-crystalline bulks [$n_{app} \geq 1$].

Category III used to be described by the KAI model [4, 5, 11]. Here we show how our model describes this Category. Let's go back to the feedback equations in Eq (12) in Ref [9] with $E_{dep}$ neglected now (see Table I) and $\alpha$ replaced by $\alpha(N)$:

$$\frac{M_0 - N - 1}{M_0 - N} = \exp\left\{-\frac{t_{N+1}}{t_\infty} \cdot \exp\left[-\frac{\alpha(N)}{E_{appl}}\right]\right\} \qquad (2)$$

The PRSPAP effect is incorporated into our model by considering a $N$-dependent $\alpha(N)$. It indicates that when $N$ (the number of switched parts) increases, the effect of PRSPAP reduces the *space-average* value of $\alpha(N)$. For given $\frac{M_0 - N - 1}{M_0 - N}$, $t_{N+1}$ decreases as $\alpha(N)$ is reduced with increasing $N$, i.e., the sample switches quicker. So it leads to a larger $n_{app}$ (e.g. $n_{app} > 1$). Therefore, if we know all the information about $\alpha(N)$ for a specific sample, which is hard, we can predict accurately its switching curve for given $E_{appl}$. [Note that according to the literature $\alpha(N, T, d, E_{appl})$ is also dependent on temperature $T$, film thickness $d$ and inversely weakly on $E_{appl}$ [5, 20, 21]; the weak dependence of $\alpha$ on $E_{appl}$ (ignored in [9] for simplicity) may lead to the large right-shift of a few time decades of the switching curves along the time axis at low $E_{appl}$, particularly for epitaxial thin films [5, 11, 17].]

For the ease of application, let us use Eq (15) instead of Eq (12) in Ref [9] for the following discussion. According to the scenario we mentioned above, we modify Eq (15) in Ref [9] to be:

$$\frac{\Delta P(t)}{2P_{M_0}} = 1 - \exp\left[-\left(\frac{t}{t_\infty}\right)^{n_{app}} \cdot \exp^{n_{app}}\left(-\frac{\alpha_{eff}}{E_{appl}}\right)\right] = 1 - \exp\left[-\left(\frac{t}{t_\infty}\right)^{n_{app}} \cdot \exp\left(-\frac{n_{app}\alpha_{eff}}{E_{appl}}\right)\right] = 1 - \exp\left[-\left(\frac{t}{\tau_{eff}}\right)^{n_{app}}\right]$$

(3)

where we introduced $n_{app}$ and effective $\alpha_{eff}$ and $\tau_{eff}$ in order to take into consideration the PRSPAP effect of *space-and-time-average* on promoting the switching process (or decreasing the switching time) [see Fig 1. Note that in reality both $n(t, space)$ and $\alpha(t, space)$ vary as a function of *time* and *location* throughout the switching process].

We notice that Eq (3) is equivalent to Eq (1) in the KAI model under the assumption of empirical Merz's law, $\frac{1}{\tau} = \frac{1}{t_\infty}\exp\left(-\frac{\alpha}{E_{appl}}\right)$. This is remarkable because of the completely different switching scenarios and approaches adopted in deriving these two theories.

**Category IV**: 90 ° domain-wall-free single-crystalline (epitaxial) thin films at medium or low $E_{appl}$ [$n_{app}$<1, ~1 or >1, depending on the dominant factor.]

Having discussed Category I and Category III, we can readily understand Category IV: recall that the promotion (or retardation) effect of $E_{dep}$ at the earlier (or later) switching stage gives rise to $n_{app}$<1 [9], and the promotion effect of PRSPAP (particularly at the middle or final switching stage) leads to $n_{app}$>1 as discussed above. Therefore, $n_{app}$ could be <1 when $E_{dep}$ dominates, ~1 when these two effects cancel each other, or >1 when PRSPAP dominates. Indeed, So *et al*. found that 1.3<$n_{app}$<1.8 for epitaxial PZT thin films of *d*=100 nm at various fields [5]. Additionally, Li and Alexe [11] reported that 1≤$n_{app}$<3 for epitaxial PZT thin films of *d*=50, 150, 200 and 215 nm containing no or few 90 ° domain walls (therefore large PRSPAP effect), and 0.4≤$n_{app}$<2.1 for the film of *d*=250 nm containing high density of 90 ° domain walls. The large number of 90 ° domain walls in the film of $d$=250 nm make the PRSPAP effect negligible with respect to the $E_{dep}$ effect at low $E_{appl}$; and this gives rise to $n_{app}$<1 as discussed previously.

Let us make a few remarks about this generalized model:

**(1) Comparing our model with the KAI model.**

From Table I, we understand that the KAI model, which neglects the $E_{dep}$ effect in thin films, can not give a good description of the polarization reversal process for the samples in Category I and Category IV. Additionally, the KAI model, which assumes unrestricted domain growth in an infinite crystal and neglects the *blocking* effect of grain boundaries and/or 90 ° domain walls in polycrystalline/ceramic samples, also makes it unable to correctly describe the switching kinetics for the samples in Category I and Category II. Therefore, one can see that the most unsuitable Category for applying the KAI model is Category I and the most suitable Category for applying it is Category III. This explains why the KAI model encounters serious problems in correctly describing the domain reversal behavior in polycrystalline thin films, particularly at medium or low applied fields (i.e., Category I), as discussed in the literature [6-9], and why it seems to be successful for single-crystalline bulks (i.e., Category III) [4] and sometimes in epitaxial thin films (i.e., Category IV when the PRSPAP effect dominates over the $E_{dep}$ effect) [5].

Note that $n_{app}$< or ~1 in Category I, II and IV seems a little unusual from the KAI model's point of view, because the KAI model implies that in ferroelectric thin films $n_{app}$=3 for the α model, and $n_{app}$=2 for the β model if we assume circular domain growth with $D$=2 (the β model assumption and $n_{app}$=2 have also been assumed for thin-film ferroelectrics by Tagantsev *et al.* [8] and Jo *et al.* [7] when they tested the validity of the KAI model). So the mixture of the α model and the β model implies that $n_{app}$ is slightly larger than 2 and lies between 2 and 3 for thin films, which is not consistent with the experimental observations ($n_{app}$ could <1, ~1 or >1) as we discussed in details in the previous sections and our previous work [9]. Similarly, for ceramic bulks the KAI model implies $n_{app}$~2-3, which obviously can not fit

Verdier's data either (see Fig 7 in Ref [9]). In Ref [9] we show that their data can be fitted correctly by our model with $n_{app}=1$.

In addition, our model uses a more realistic $E_{tot}$-activated nucleation and statistical switching scenario for all the parts of a ferroelectric capacitor [9], which is also new. That is because in the KAI model Ishibashi assumed two scenarios for nucleation without detailed justifications: the *α* model, where nucleation occurs with a constant rate during switching, and the *β* model, where nucleation occurs only at $t=0$ [1]; while in the NLS model Tagantsev assumed a broad mesalike distribution function for logarithmic nucleation waiting time ($\log \tau$) instead [8].

Furthermore, the KAI model was built by assuming a field-independent constant "domain velocity $v$" for Category III, which might also be problematic. In contrast, our model doesn't rely on the concept of constant "domain velocity" or continuous DWM (see Fig 1). Even for the samples falling into Category III and IV, the generalized version of our model indicates that the *local* activation field $\alpha_{loc}(N)$ of one adjacent non-switched part generally decreases and sometimes even slightly increases depending on the local geometry of the adjacent switched domain parts and the local defect structure (Fig 1). Therefore, DWM or "apparent" DWM [22] in our model is *abrupt* or *discontinuous*, and contains a lot of "leaps". According to Eq (2), it implies that the *local* "domain velocity $v_{loc}$" varies significantly depending on the local domain geometry, which is consistent with the observations in the literature [23]. Therefore, the assumption of constant "domain velocity $v$" or continuous DWM in the KAI model is indeed questionable.

The KAI model under the β model assumption predict that $1/\tau \sim v$ [1]. Considering Merz's law and Eq (3), we have:

$$v_{eff} \sim \frac{1}{t_\infty} \cdot \exp\left(-\frac{\alpha_{eff}}{E_{appl}}\right) \sim v_\infty \cdot \exp\left(-\frac{\alpha_{eff}}{E_{appl}}\right) \qquad (4)$$

where the effective or apparent "domain velocity $v_{eff}$" is defined as the *space-average* value of all the $v_{loc}$ around one cluster of switched domain parts at given time point. The relationship of this type has been experimentally confirmed and widely used in the literature [16, 18, 24, 25]. In Ref [16, 18], the authors showed that DWM in epitaxial ferroelectric thin films is a creep process and follows a modified formula, $v \sim \exp\left[-\left(\dfrac{\alpha}{E_{appl}}\right)^{\mu}\right]$, where $\mu$ is a dynamic exponent and $\mu=1\pm0.2$ [16, 18].

**(2) Switching current peak and the influence of pulse rise time and *RC* constant of the circuit**

The traditional approach to measure the polarization reversal of a ferroelectric capacitor is to measure the switching current directly, not the pulse method shown in the inset of Fig 3 in Ref [9]. Our model (Eq (3)) [or the KAI model, Eq (1)] predicts the switching current density $i_{sw}(t)$ to be:

$$i_{sw}(t) = \frac{dP(t)}{dt} = \frac{2P_{M_0} n_{app}}{\tau_{eff}} \left(\frac{t}{\tau_{eff}}\right)^{n_{app}-1} \cdot \exp\left[-\left(\frac{t}{\tau_{eff}}\right)^{n_{app}}\right] \qquad (5)$$

Fig 2 shows the switching-current profiles for $n_{app}$=3, 2, and 1 according to Eq (5) and Merz's law, where $E_{appl}$=150 kV/cm, $P_{M_0}$=30 µC/cm$^2$, $\alpha$=500 kV/cm and $t_\infty$=1 ns ($E_{dep}$ is neglected). The switching current curve for $E_{tot} = E_{appl}+E_{dep} =$150 kV/cm+$E_{dep}$ corresponding to $n_{app}$<1 (See Fig 5 in Ref [9], where $d$=200 nm, $\varepsilon_i/d_i$=20 nm$^{-1}$) has also been plotted here for comparison. One can see that switching peaks only occur for $n_{app}$>1. For $n_{app}$=1 and $n_{app}$<1, the curves show a plateau at low $t$, followed by a logarithmic decay of switching current. We also see that the smaller $n_{app}$ the more decades the switching current expands. We notice that for most of the samples falling in Category I, II and IV there is essentially no switching current peak that could be observed as confirmed experimentally [6, 26, 27]. The reason why sometimes people did see these peaks in polycrystalline thin films is probably due to either the PRSPAP effect (which increases $n_{app}$ to $n_{app}$>1), or the contamination of the $i_{sw}(t)$ by the *extrinsic* capacitor charging curve/peak induced by the combination of pulse rise time (which varies from 50 ps [26] to 10 ns [8]) and *RC* constant (which varies from 45 ps [26] to ~100 ns [28]).

In summary, the switching kinetics in ferroelectrics including both epitaxial/polycrystalline thin films and single-crystalline/ceramic bulks at various applied fields has been classified into four categories, depending on whether the depolarization field and/or the effect of PRSPAP can be neglected. We show that our statistical model developed recently in its generalized version applies to all these four categories. Finally, the comparison between our model and the conventional KAI model has been made and the behavior of switching currents for different *n* has been discussed.

Table I: four switching categories for the switching kinetics of ferroelectrics

| **The present model** | $E_{dep}$ has to be considered | $E_{dep}$ can be neglected |
|---|---|---|
| PRSPAP has to be considered | Category IV: *90 º domain-wall-free single-crystalline (epitaxial) thin films at medium or low $E_{appl}$* [$n_{app}$<1, ~1 or >1, it depends on the dominant factor.] | Categorys III *90 º domain-wall-free single-crystalline (epitaxial) thin films at high $E_{appl}$ or single-crystalline bulks* [$n_{app} \geq 1$] |
| PRSPAP can be neglected | Category I: *Polycrystalline thin films (or relatively thicker epitaxial films containing blocking 90 º domain walls) at medium or low $E_{appl}$* [$n_{app}$<1] | Category II: *Polycrystalline thin films (or relatively thicker epitaxial films containing blocking 90 º domain walls) at high $E_{appl}$ or polycrystalline (ceramic) bulks* [$n_{app}$~1] |

**Figure Captions:**

Fig 1. (color online) Schematic diagram of polarization reversal scenario in the present model after incorporating the PRSPAP effect. The shaded (green) regions denote the switched parts with spontaneous polarization pointing downwards, while the white regions denote the non-switched (or retained) parts with spontaneous polarization pointing upwards. In any cluster of switched parts, the "darkest green" region represents the one that switched earliest and therefore did not rely on the switching promotion/DWGN effect of the adjacent parts, while the "lighter-green" regions represent the ones that switched later and therefore did rely on the switching promotion/DWGN effect of the adjacent switched parts. Note that the figure is not to scale.

Fig 2. (color online) the switching-current profiles for $n_{app}$=3, 2, and 1 according to Eq (5). Also shown is the switching current curve for $E_{tot}= E_{appl}+E_{dep}$ =150 kV/cm+$E_{dep}$ corresponding to $n_{app}$<1.

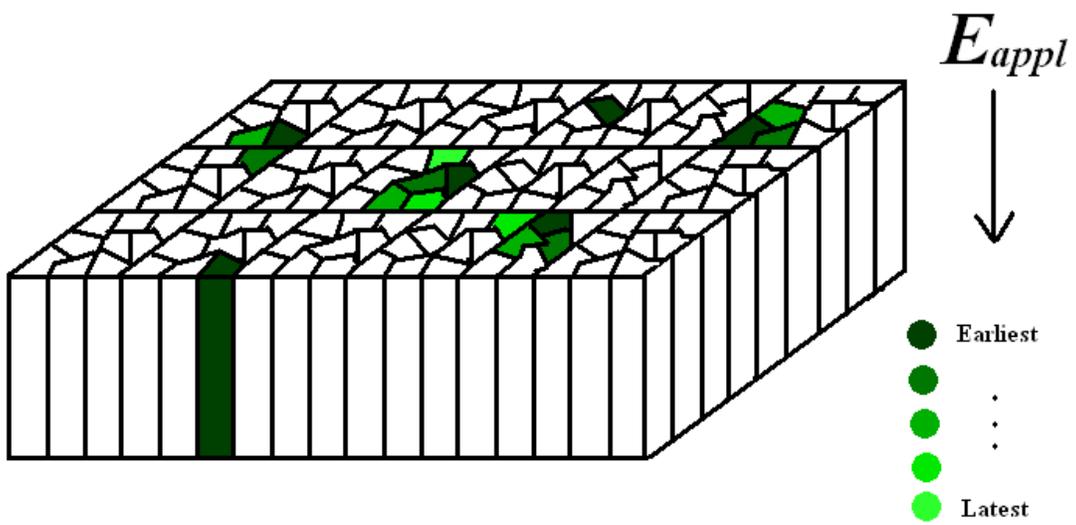

Fig 1

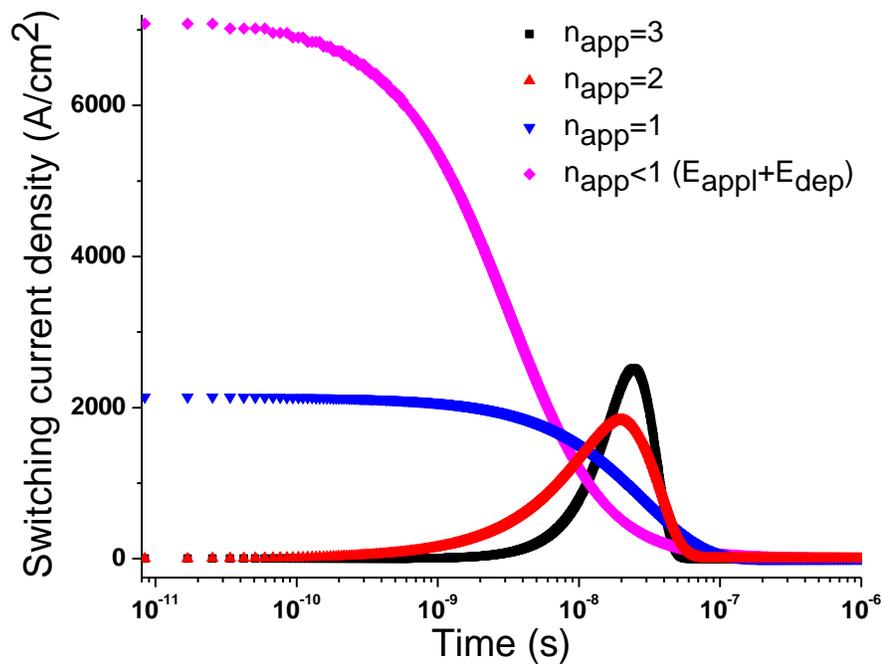

Fig 2